\documentclass[letterpaper]{JHEP3}
\usepackage{graphicx}
\usepackage{epsfig}

\title{Boundary Localized Symmetry Breaking and Topological Defects}
\author{R Holman \\ Department of Physics\\ Carnegie Mellon University\\ Pittsburgh, PA 15213 USA\\ E-mail: \email{rh4a@andrew.cmu.edu}}
\author{Matthew R Martin\\ T-8, MS B285\\ Los Alamos National Laboratory\\ Los Alamos, NM 87545 USA\\ E-mail: \email{mrmartin@lanl.gov}}

\abstract{We discuss the structure of topological defects in the context of extra dimensions where the symmetry breaking terms are localized.  These defects develop structure in the extra dimension which differs from the case where symmetry breaking is not localized. This new structure can lead to corrections to the mass scale of the defects which is {\em not} captured by the effective theory obtained by integrating out the extra dimension.  We also consider the Higgsless model of symmetry breaking and show that no finite energy defects appear in some situations where they might have been expected. }

\preprint{LA-UR-05-1366\\ CMU-HEP-05-01}

\begin{document}

\section{Introduction}
Much recent model building work has made use of additional dimensions beyond the 3+1 with which we are familiar.  These efforts are largely motivated by an attempt to relate the vastly different scales of gravity and the electroweak theory in a natural way.  The possibilities for doing this have been greatly expanded through many new tools which become available with extra dimensions, but one common feature to nearly all of these models is the presence of symmetry breaking terms localized to a hyper-surface (brane) in the extra dimension.  For some models, it is simply the electroweak symmetry which is broken at a boundary~\cite{Davoudiasl:1999tf, Huber:2000ie, Arkani-Hamed:2001mi, Csaki:2002gy}, while in other cases it could be left-right symmetry~\cite{Higgsless,Mohapatra:1999pj, Mimura:2002te} or GUT symmetry~\cite{Pomarol:2000hp, Chaichian:2002uy, Goldberger:2002pc, Kyae:2002ss, Carena:2003fx, Agashe:2004bm} that is broken on the brane. 

The formation and evolution of topological defects which result from the breaking of some symmetries is also a well studied topic.  The type of defect that could form in a symmetry breaking transition depends on the topology, specifically the homotopy structure of the vacuum manifold of the theory, while how many form in a given situation as well as their subsequent evolution is a dynamical question~\cite{kibble,zurek}. It is well known, for example that GUT symmetry breaking allows for the formation of monopoles and that they can give rise to cosmological problems~\cite{monopole}.  In addition, many models lead to either global or local strings and walls, as can be found in the comprehensive reviews~\cite{VilenkinShellard,VilenkinReview,Hindmarsh:1994re}.  Even the electroweak symmetry breaking in the standard model allows for textures which separate different vacua labeled by the Higgs field winding number or the Chern-Simons number.  Transitions between these vacua mediate changes in baryon plus lepton number and are a key ingredient in models of electroweak baryogenesis~\cite{troddenrev}.

In the presence of an extra dimension, defect solutions have been extended from the 4 dimensional case \emph{if} the extra dimension is homogeneous~\cite{HomoExtraD}.  However, when the symmetry breaking is localized so that the symmetry remains unbroken in the bulk of the extra dimension, the structure of these defects may be modified. The essential reason for this modification can be understood in a simple way.  When the symmetry is broken throughout the bulk of the extra dimension, the field profile for defect solutions is homogeneous in that extra dimension because the potential is homogeneous.  In some sense, the defect itself is spread out across the new dimension.  However, if the symmetry breaking is localized to one boundary, then the defect will try to interpolate between the symmetry breaking solution on the boundary and the unbroken solution in the bulk.  It is now possible that the defect will in some sense (to be made more precise later) be localized to the symmetry breaking brane.  It is also possible that the defect will remain, more or less, spread across the extra dimension. Which situation is actually realized will depend in general on the type of symmetry under consideration as well as on the choices of  parameters in the scalar potential.

In this paper, we consider the structure of topological defects in spaces with one flat extra dimension where the symmetry breaking is localized on a single brane. To be specific,  we will look at the static solutions for both a global and local U(1) symmetry.  In 3+1 infinite dimensions it is well known that these models lead to global and local strings, respectively.  Similarly, in our extra dimensional setup defects will form for some values of parameters, {\em even if} the symmetry is not broken in the bulk of the extra dimension.  We will show that this symmetry breaking can be understood from the perspective of a simple 4 dimensional effective theory, but also that the effective theory does {\em not} capture enough of the 5 dimensional physics to correctly predict the tension of the resulting strings.  This is actually equivalent to the statement that solutions which are homogeneous in the bulk do not correctly describe the strings coming from boundary localized symmetry breaking.

Of course if the defects are spread across the extra dimension in any sense then it is no longer strictly correct to call them strings.  However, from the perspective of the low energy observer who can only probe the large $3+1$ dimensions, they still look like strings.  For lack of a better term, we will continue to call all these structures strings in both the 4 and 5 dimensional theories. Similar situations can occur for other defects as well. 

In the case of a {\em global} symmetry, the localization of the symmetry breaking induces a non-trivial profile for the scalar field, especially near the core of the defect.  In the case of a {\em local} $U(1)$ symmetry, only the gauge field has a non-trivial profile in the extra dimension. For either case, it is clear that the 4-dimensional effective field theory obtained by na\"{i}vely integrating out the extra dimension will not contain this information. In particular, observers restricted to the symmetry breaking brane will not be able to infer the correct value of the string tension $\mu$. This could have some cosmological applications since the evolution of a string network depends crucially on the dimensionless parameter $G_N\mu$ \cite{kibblehindmarsh}. 

Similarly, in the electroweak theory, the energy scale associated with the sphaleron saddle-point configuration between two vacua controls the rate of baryon violating transitions in models of electroweak baryogenesis~\cite{troddenrev}.  What this means is that, when looking at defects in higher dimensions, care must be taken to incorporate the {\em full} defect profile, including its behavior in the extra dimensions in order to compute the correct value of the defect's energy density. 

In the case of the local symmetry, another question arises when we consider the Higgsless models of electroweak symmetry breaking~\cite{Higgsless}.  In this category of models, the symmetry is spontaneously broken through the choice of boundary conditions rather than by placing a Higgs field on that boundary.  However, these same boundary conditions may be realized by making use of the Higgs mechanism and then taking the Higgs vacuum expectation value (VEV) to infinity.  By repeating this procedure with a defect solution, we will see that an Abelian theory with a broken local symmetry will only support infinitely massive and small topological defects in the absence of a Higgs field.  We make use of a generalization of Derrick's theorem\cite{derrick} to show that in three large dimensions no finite energy static defects may exist without a Higgs field.  Similarly, non-Abelian theories are somewhat constrained.  The Higgsless models may then provide another possible method for removing unwanted defects from a theory.

Section~\ref{sec:global} contains a discussion of the breaking of a global $U(1)$ symmetry which is broken in to two parts.  We first briefly discuss the 4 dimensional case which, while well known, is necessary to make accurate comparisons to the extra dimensional case in the second part.  We then discuss the breaking of a local $U(1)$ symmetry in section~\ref{sec:local} and again compare the 4 and 5 dimensional solutions, taking a special interest in comparing the string tensions as a function of the size of the extra dimension.  Finally, section~\ref{HiggslessSection} treats the Higgsless models and we summarize our results in section~\ref{SummarySection}.

\section{Global Symmetry Breaking}\label{sec:global}
We wish to begin consideration of localized symmetry breaking with a very simple example where it is possible to visualize the solutions easily and where some aspects of the problem can be solved analytically.  We will therefore consider a single scalar field with a global $U(1)$ symmetry.  After symmetry breaking the vacuum manifold is topologically a circle which, through winding at infinity, can support the well known global string in three spatial dimensions~\cite{VilenkinShellard}.  In the presence of an extra dimension where the vacuum manifold is still a circle, but the symmetry breaking is localized, we will find a modified string with a new structure along the extra dimension and a modified tension.

\subsection{Global Strings in Four Large Dimensions.}\label{subsec:usual}
In preparation for comparison with the extra dimensional case, we first review the standard breaking of a global $U(1)$ symmetry in $3+1$ large dimensions.  We will want the results from this case to understand what new features arise from the addition of an extra dimension, as well as to make a comparison of the energy density in the 4 and 5-dimensional strings.  We start with the following action for a complex scalar field:
\begin{equation}
S = \int d^{4}x \left[\eta^{\mu\nu}\partial_\mu \phi^* \partial_\nu \phi - {\lambda_4 \over 4}
\left(|\phi|^2 -v_4^2\right)^2\right], \label{action4d}
\end{equation}
where the scalar has a constant, homogeneous solution, $|\phi| = v_4$, with zero energy density.  The static solutions which correspond to a topological defect have a winding of the form
\begin{equation}
\phi(r,\theta,z) = v_4 e^{i n \theta} f(r),
\end{equation}
where $r$ is the radius from the core of the string and $\theta$ is the polar coordinate going around the string.  These solutions are both homogeneous in the third spatial direction, $z$, as well as static.
From the action~(\ref{action4d}) we can derive the equation of motion for $f(r)$:
\begin{equation}
-{1 \over r} \partial_r \left(r f'(r)\right) + {n^2 \over r^2}f(r) 
+ {\lambda_4 v_4^2 \over 2}\left(f(r)^2 - 1\right)f(r) =0,
\end{equation}
which we will solve numerically later.  It would be reasonable to simplify this equation further by rescaling the radial coordinate by $v_4$ to make it dimensionless and remove $v_4$ from the equation.  However, to make comparison with the 5 dimensional case later, we leave the $v_4$ explicit.  These winding solutions must reduce to the vacuum far from the core of the string and have continuous field values in the core, so $f(r)$ has the boundary conditions:
\begin{equation}
f(0)=0, \qquad \lim_{r \to \infty} f(r) = 1.
\label{bc4d}
\end{equation}

Once we have a solution for $f(r)$ we may calculate the tension of the string by integrating the energy density over the two spatial dimensions transverse to the string:
\begin{eqnarray}
\mu_4 &=& \int r\, dr \, d\theta \, \rho(r,\theta) \nonumber \\
&=& 2 \pi v_4^2 \int r\, dr \left[ f'^2 + {n^2 \over r^2} f^2
+ {\lambda_4 v_4^2 \over 4} \left( f^2 - 1\right)^2\right].
\end{eqnarray}
It is well known that this integral for the global string tension does not converge.  This may be seen through a power series expansion far from the string showing that the energy density does not fall to zero fast enough.  If we suppose that there is a network of strings with positive and negative winding number separated by some characteristic scale $R$ then we may impose a large distance cutoff on the integral.  In this case the string tension scales with the log of the cutoff scale:
\begin{equation}
\mu_4 \sim 2\pi v_4^2 n^2 \ln (v_4 R).
\end{equation}

\subsection{Symmetry Breaking with a Homogeneous Extra Dimension}\label{subsec:homextradim}
If we imagine for a moment, an extra dimension compactified on a circle (with periodic boundary conditions), such that the theory is homogeneous in the extra dimension, then the above work extends trivially.  The winding solution will simply be homogeneous in the new dimension.  In other words, this is a 2-brane which wraps around the compact 5th dimension.  From the perspective of a low energy observer who sees only the large uncompactified dimensions, this looks like a string as we would expect.

We can illustrate the point by extending the action above to a fifth dimension $y$:
\begin{equation}
S = \int d^{4}x \int_0^L dy \left[{\eta^{\mu\nu}\partial_\mu \phi^* \partial_\nu \phi \over L} -
{\lambda_4 \over 4L} \left(|\phi|^2 - v_4^2\right)^2\right],
\end{equation}
where the factors of $L$ are chosen such that engineering dimensions of the coefficients and $\phi$ are the same as the 4 dimensional case.  The solution, as above is:
\begin{equation}
\phi(r,\theta,z,y) = v_4 e^{i n \theta} f(r),
\end{equation}
with homogeneity along the string and in the extra dimension (in $z$ and $y$ respectively).  The function $f$ must satisfy the same equation of motion and boundary conditions as before.  We may introduce an effective theory which captures all of the physics of the string by replacing $\phi(x^\mu,z) \to \phi(x^\mu)$ and integrating out the extra dimension.  The point here is that this procedure for writing down the effective theory is the same whether we are considering vacuum solutions or winding solutions.  The reason is that neither the vacuum nor the string solutions depend on the extra coordinate, $z$.  Doing this yields an effective action that is the same as the original 4-dimensional action.  By contrast, if a solution with non-trivial winding were to have a structure along the extra dimension which differs from the vacuum solution then we would need a different effective theory for each value of the winding number $n$.  This is what we will find in the next sections.

\subsection{Symmetry Breaking Localized on a Brane}\label{subsec:localizedssb}
Suppose now that the  symmetry breaking terms are localized in the extra dimension.  The bulk contains only a mass term for our complex scalar, while the brane has the usual symmetry breaking term:
\begin{equation}
S = \int d^4x \int_0^L dy \left\{ \eta^{MN} \partial_M \phi^* \partial_N \phi - m^2 |\phi|^2
- \delta(y-L) {\lambda \over 4 \Lambda^2} \left(|\phi|^2 - v^3 \right)^2 \right\}. \label{globalU1action}
\end{equation}
Again, $y$ labels the extra compact dimension.  The notation is chosen so the $\lambda$ is dimensionless, while $v$, $\Lambda$, and $m$ all have units of mass.  The metric is mostly minus with $M$ and $N$ running over all five dimensions.  The important change from the case of the homogeneous extra dimension is that the vacuum solution is necessarily dependent on $y$.  In the bulk, the potential is minimized at $\phi=0$, but on the symmetry breaking brane at $y=L$, the potential is minimized by $|\phi| = v^{3/2}$.  Any solution will therefore interpolate between these two and the VEV will be something less than that set by the scale $v$ on the brane.  It is easy to find a solution which is homogeneous in the 3 infinite dimensions, but the winding solution will require a numerical calculation.

The equation of motion resulting from the action of eq.~(\ref{globalU1action}) is:
\begin{equation}
\partial_t^2\phi - \vec{\nabla}^2\phi - \partial_y^2\phi + m^2\phi = 0.
\end{equation}
We explicitly separate out the time and space components of the derivative because we will be seeking static solutions.  The vacuum solution is
\begin{equation}
\phi = a \cosh(m y) + b\sinh(m y),
\end{equation}
where we have used the symmetry to make $\phi$ real.  We will sometimes refer to this solution as the $n=0$ solution since there is no winding.  The boundary conditions also come from the variation of the action, taking care with the integration by parts.  They are:
\begin{eqnarray}
\partial_y\phi|_{y=0} &=& 0 \nonumber \\
\partial_y\phi|_{y=L} &=& -{\lambda \over 2 \Lambda^2} \left(|\phi|^2 - v^3\right)\phi|_{y=L}.
\end{eqnarray}
These conditions imply $b=0$ and
\begin{equation}
a = {1 \over \cosh(m L)} \sqrt{v^3 - {2 m \Lambda^2 \over \lambda}\tanh(m L)}.
\label{aVEV}
\end{equation}
Na\"{i}vely, we could allow the imaginary solution for $a$,  since the field is complex. However, the boundary condition cannot be satisfied if $v^3 < 2m\Lambda^2/\lambda \tanh(mL)$ since for small enough $v$, the positive bulk mass squared term dominates the negative brane symmetry breaking term and the only solution is $\phi \equiv 0$.  We will see this more clearly from the effective theory in the next section.

We may calculate the effective 4 dimensional energy density of this solution by integrating over the extra dimension:
\begin{eqnarray}
\rho^{(4)}_{n=0} &=& \int_0^L dy \left[ \partial_y\phi^2 + m^2\phi^2 
+\delta(y-L) {\lambda \over 4 \Lambda^2} \left(\phi^2-v^3\right)^2 \right] \nonumber \\
&=& m \tanh(m L) \left(v^3 - {m \Lambda^2 \over \lambda} \tanh(m L) \right).
\end{eqnarray}
This is the (classical) cosmological constant in this model which we will need to subtract off of the energy density for the winding solutions in order to compare with the 4 dimensional theory where the cosmological constant was already set to zero.
That this solution is in fact lower in energy than the trivial $\phi \equiv 0$ solution can be seen from the difference in 4 dimensional energy densities:
\begin{eqnarray}
\rho^{(4)}_{\phi=0} - \rho^{(4)}_{n=0} = {\lambda v^6 \over 4 \Lambda^2}
\left(1 - {2 m \Lambda^2 \over \lambda v^3}\tanh(mL) \right)^2 > 0.
\label{global_energy_difference}
\end{eqnarray}

\subsection{The Four Dimensional Effective Action}\label{subsec:4dEffAct}
We have seen that the VEV of the field on the symmetry breaking brane is not set by $v$ alone as it would be if the bulk did not exist or if $m=0$.  Instead it is
\begin{equation}
\phi(y=L) = \sqrt{v^3 - {2 m \Lambda^2 \over \lambda}\tanh(mL)} < v^{3/2}.
\end{equation}
This can be understood in a quantitative manner from the perspective of a 4 dimensional effective theory by promoting $a$ to a 4 dimensional scalar.  We make the replacement
\begin{equation}
\phi \rightarrow a\left(x^{\mu}\right) \cosh(m y)
\end{equation}
in the action~(\ref{globalU1action}) and integrate over the bulk.
The resulting action does not yet have a canonically normalized field, but we can still consider the location of the minimum by setting the first derivative of the potential to zero.  It is located exactly at the value of $a$ given in equation~(\ref{aVEV}) by solving the bulk equation of motion.

Once we do rescale the field $a(x^{\mu})$ we may compare with the action in equation~(\ref{action4d}) and identify the effective 4-dimensional mass squared and coupling, $-\lambda_4 v_4^2$, $\lambda_4$:
\begin{eqnarray}
\lambda_4 &=& {4 \lambda \over \Lambda^2 L^2} {\cosh^4(mL) \over
\left(1 + {\sinh(2mL) \over 2mL} \right)^2} \label{CouplingComparison} \\
m_4^2 = -\lambda_4 v_4^2 &=& {2 \lambda \over L \Lambda^2 }
{\cosh^2(mL) \over 1+ {\sinh(2mL)\over 2mL}}
\left(-v^3 + {2 m \Lambda^2 \over \lambda} \tanh(mL) \right). \label{MassComparison}
\end{eqnarray}
Note that the 4-d effective mass squared is negative for exactly the region of parameter space which allowed a non-zero value for $a$ in the previous section.  Decreasing $v$ (or increasing the bulk mass) just increases the mass squared in the 4-d theory until the symmetry is restored.  These relations will be important later to compare the 4-dimensional and brane world calculations of the string tension.

\subsection{Non-Trivial Winding}\label{subsec:nontrivwinding}
We now want to consider solutions which wind on the brane.  These solutions will necessarily depend on both $y$ and $r$.  The delta function contribution to the potential on the brane means that the solution must have $y$ dependence, interpolating again between $\phi=0$ in the bulk and $\phi=v^{3/2}$ on the brane.  The winding of these solutions further implies that the field must go to zero in the core of the string and approach the vacuum solution far from the string, thus having $r$ dependence as well.  Separable solutions for the differential equation will not work because of the nonlinear boundary condition on the symmetry breaking brane, so we try the following anzatz:
\begin{equation}
\phi = \rho(r,y)e^{i n \theta},
\end{equation}
with $\rho$ real.  Again, the solution we are seeking is homogeneous along the string and static.  As in the 4 dimensional case without a brane, we must impose the boundary condition $\rho=0$ at $r=0$ to render the field continuous.  Also, infinitely far from the string, the field should approach the value of the $n=0$ solution.
\begin{equation}
\lim_{r \to \infty} \rho(r,y) = a \cosh(m y).
\end{equation}
The boundary conditions at $y=0$ and $y=L$ are unmodified from above.  With these four boundary conditions and the second order bulk equation of motion, the problem is completely specified and we may numerically solve for the profile of this ``string'' (recall our comments above; it really is a 2-brane, but will look like a string to observers restricted to the symmetry breaking brane).

Before this problem can be solved numerically, we make the equations dimensionless by scaling by the appropriate power of the bulk mass $m$, and by performing the rescaling $\rho \to v^{3/2} \rho$.  The bulk equation of motion now reads:
\begin{equation}
\left( \partial_y^2 + {1 \over r} \partial_r (r\partial_r) - {n^2 \over r^2} - 1 \right) \rho(r,y)=0,
\end{equation}
while the four boundary conditions are:
\begin{eqnarray}
\partial_y \rho(r,y)|_0 &=& 0 \\
\partial_y \rho(r,y)|_L &=& -\tilde{\lambda} \left(\rho^2(r,L)-1\right)\rho(r,L) \\
\rho(r=0,y) &=& 0 \\
\lim_{r \to \infty} \rho(r,y) &=& {\cosh(y) \over \cosh(L)} \sqrt{1- \tanh(L)/\tilde{\lambda}}. \label{bc4}
\end{eqnarray}
Only three dimensionless parameters are left: $L$, $n$, and
\begin{equation}
\tilde{\lambda} \equiv {\lambda v^3 \over 2 \Lambda^2}.
\end{equation}

Starting from an appropriate {\em ansatz} satisfying the boundary conditions, we may use a relaxational technique to solve for the profile of $\rho(r,y)$.
As expected, the system settles down into the solution we are seeking and on the symmetry breaking brane (at $y=L$) has the profile given in figure~\ref{phi_winding_profile}.
\DOUBLEFIGURE[t]{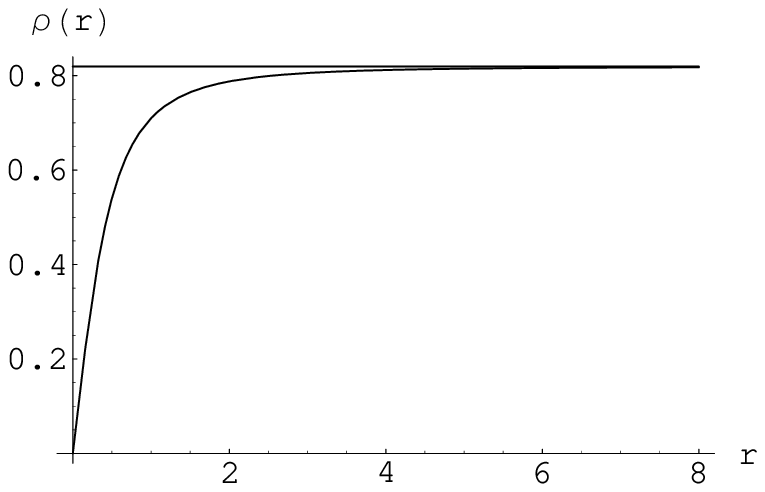,width=.4\textwidth}
{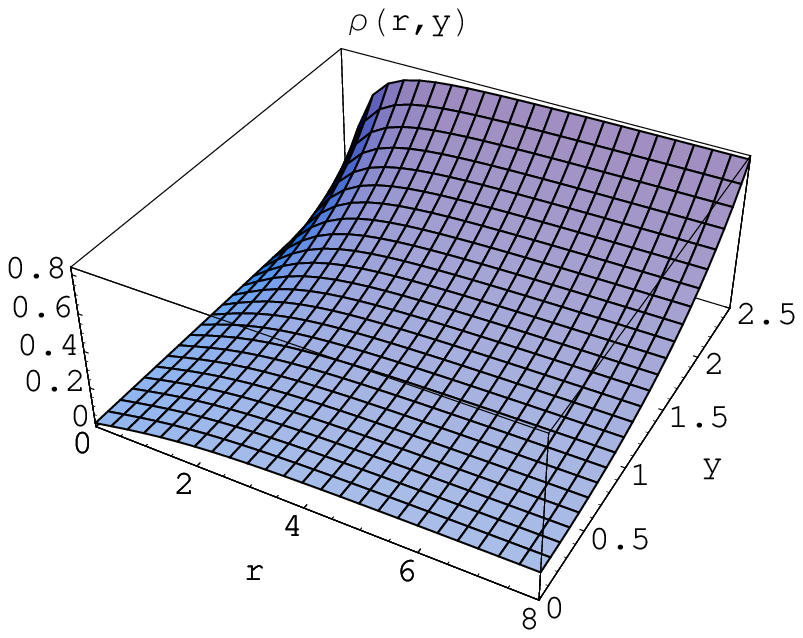,width=.4\textwidth}
{The value of the string profile on the symmetry breaking brane, $\rho(r,L)$, for the two solutions: $n=0$ (flat line) and $n=1$.  All quantities are in units of the bulk mass parameter.  We have chosen $n=1$, $L=5/2$, and $\tilde\lambda = 3$. \label{phi_winding_profile}}
{The value of $\rho(r,y)$ throughout the bulk with the same parameters as before.  The value for $r$ goes from $0$ to $8$, while $y$ goes from $0$ to $2.5$. \label{phi_winding_bulk_profile}}
More generally, the solution for $\rho(r,y)$ throughout the bulk is shown in figure~\ref{phi_winding_bulk_profile}.
We may numerically integrate the 4-dimensional energy density over $r$ and $\theta$ to get the mass per unit length of this defect.  Of course there will be a divergent contribution coming from the fact that the homogeneous, $n=0$, solution has non-zero 4-dimensional energy density (cosmological constant).  So, instead we calculate the difference in energy between these two solutions to get the mass per unit length of string:
\begin{eqnarray}
\mu &=& 2\pi v^3 \int r dr \bigg[
\int dy \left( \partial_y \rho^2 + \partial_r \rho^2 + {n^2 \over r^2}\rho^2 + \rho^2 \right) \nonumber \\
&& \quad \qquad \qquad +{ \tilde{\lambda} \over 2} \left(\rho(r,L)^2-1\right)^2 
- \tanh(L) \left(1-{1 \over 2 \tilde\lambda}\tanh(L)\right) \bigg].
\end{eqnarray}
Notice that although $v$ is no longer a part of the equations of motion or boundary conditions, it still affects the string tension in a way similar to the case of the 4-dimensional string.  Recall that in this expression $v$ is dimensionless and this string tension is in units of $m$.  We plot the 4-dimensional energy density as a function of $r$ in figure~\ref{string_tension_comparison} (that is the quantity inside the square brackets in the expression for the string tension).  As in the case of the 4-dimensional global string, the string tension diverges logarithmically with radius.  In fact, since the winding solution approaches the non-winding solution far from the string and the physics of the non-winding solution is well described by the effective theory obtained by integrating out the extra dimension, this agreement in energy density for the winding of the full theory and the effective theory far from the string is expected.

We now turn to the 4 dimensional problem specified in section~\ref{subsec:usual}.
In order to make a direct comparison between the effective and full theories, we need the relation between the 4 and 5 dimensional parameters in the rescaled, dimensionless units.  Equations~(\ref{CouplingComparison}) and~(\ref{MassComparison}) become:
\begin{eqnarray}
\lambda_4 &=& {8 \tilde\lambda \over v^3 L^2} {\cosh^4(L) \over
\left(1 + {\sinh(2L) \over 2L} \right)^2} \label{fourDlambda} \\
\lambda_4 v_4^2 &=& {4 \tilde\lambda \over L } {\cosh^2(L) \over 1+ {\sinh(2L)\over 2L}}
\left(1 - {1 \over \tilde\lambda} \tanh(L) \right). \label{fourDvev}
\end{eqnarray}
Notice that $v_4^2$ is proportional to $v^3$ so the tension for both the 4 and 5 dimensional strings have a leading factor of $2\pi v^3$ which can be dropped.  We then calculate the energy per unit length of string as a function of $r$, the distance from the string.  This is shown in figure~\ref{string_tension_comparison}.  The shape of both curves are similar, however, the amplitudes near the core of the string differ: the energy density for the effective theory is larger.
\FIGURE[t]{\epsfig{file=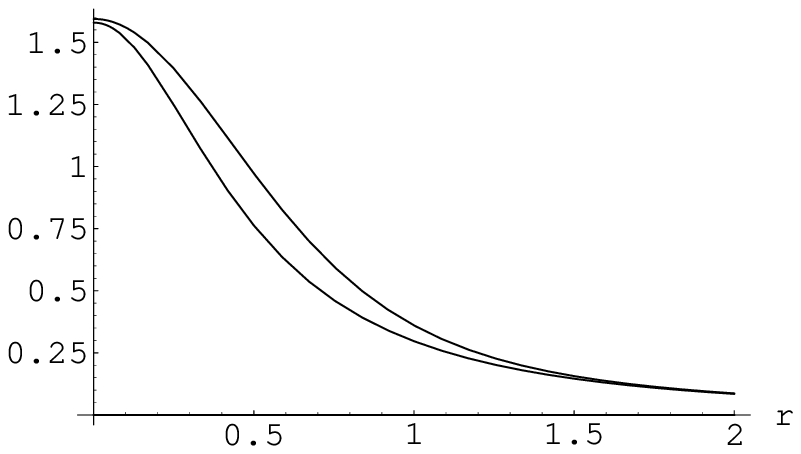,width=.6\textwidth}
\caption{A comparison of the energy density as a function of $r$ in the cases where symmetry breaking happens on the brane (lower curve) and the pure 4-d model (upper curve).  The string tension comes from multiplying these curves by $2\pi v^3 r$ and integrating over $r$. The difference goes to zero faster than $1/r^2$ at large $r$, so they have identical large distance log divergences as they must if the effective theory is to capture the IR physics.}
\label{string_tension_comparison}}
To compare the actual string tension, we need to impose a cutoff in $r$, which in the absence of a cosmological context for our problem is fairly arbitrary.  By integrating to $r=2$, we find the relative difference in the string tensions to be $17 \%$.  Of course choosing a much larger distance cutoff would make the relative difference in tension smaller.

\subsection{Limiting Behavior}
Several regions of parameter space can be understood analytically.  We have already seen that for large values of the bulk mass or small values of $v$, the symmetry is restored and there are no winding solutions.  In terms of the dimensionless parameters, this corresponds to small $\tilde{\lambda}$.  For large $\tilde{\lambda}$, the field on the symmetry breaking brane far from the string will approach $v^{3/2}$ as can be seen simply from the boundary condition~(\ref{bc4}).  In this case, the potential on the boundary is dominating over the mass term in the bulk.

One check we can make is to consider the small $L$ limit.  In this case we expect that the effective theory should give a very good description of the physics of the string in the full theory.  That this is the case can be seen from expanding the solution for $\rho$ in $m\,y$:
\begin{equation}
\rho(r,y) = A(r) + B(r)(m\,y)^2 +\ldots,
\end{equation}
where the term linear in $y$ must have coefficient zero to satisfy the boundary condition at $y=0$.  As may be expected, $B(r)$ is order one as seen in numerical solutions, and so for a small extra dimension ($mL \ll 1$) even the winding solution is nearly homogeneous across the extra dimension.
Using this expansion for the wave function, integrating over the $y$ direction, and then rescaling the scalar field by $\sqrt{L}$, the action in equation~(\ref{globalU1action}) reproduces the 4 dimensional theory to leading order in $mL$
\begin{eqnarray}
S &=& \int d^4x \left\{ \partial_\mu\phi^* \partial^\mu\phi
-{\lambda \over 4 \Lambda^2 L^2} \left( |\phi|^2 - v^3L \right)^2\right\} \left(1 + {\cal O}(mL)\right).
\end{eqnarray}
This is just a restatement of the fact that for a homogeneous theory, the effective theory captured all of the physics of the string.  It may be checked that the parameters in this action agree with the small $L$ limit of the 4-dimensional parameters in equations~(\ref{CouplingComparison}) and~(\ref{MassComparison}).

We may also consider the large $L$ limit.  Far from the string, the field value drops off exponentially away from the symmetry breaking brane.  So although this object is extended in the extra dimension, its profile is suppressed.  On the symmetry respecting brane and far from the string, the field approaches:
\begin{equation}
\phi = \sqrt{v^3-\frac{2m\Lambda^2}{\lambda}} e^{in\theta} e^{-mL}.
\end{equation}
So in this sense the defect is trapped on the symmetry breaking brane.
Of course, the masses of KK excitations decrease as the size of the extra dimension grows, and so in any phenomenological context there would be a limit on how far this parameter could be pushed.

\section{The Local String}\label{sec:local}
We now turn to consider the breaking of a local $U(1)$ symmetry by a boundary localized Higgs field.  This problem is more interesting from the point of view of model building since it is the Abelian version of some models of physics beyond the standard model and because we will be able to explore the interesting question of whether topologically stable defects may form in the absence of a Higgs field when we take the large VEV limit.  We begin again with the 4 dimensional theory, which has been discussed at length~\cite{VilenkinShellard}, so that we may clearly contrast this with the extra dimensional model.

\subsection{The Four Dimensional Local String}
We start with the action
\begin{eqnarray}
S = \int d^4x \left[ \left( {\cal D}_{\mu}\phi \right)^* {\cal D}^{\mu}\phi
-{1 \over 4} F_{\mu\nu}F^{\mu\nu} 
- {\lambda \over 4} \left( |\phi|^2 - v^2\right)^2 \right],
\end{eqnarray}
where
\begin{eqnarray}
{\cal D}_\mu \phi &=& \left(\partial_\mu - i e A_\mu \right) \phi \\
F_{\mu\nu} &=& \partial_\mu A_\nu  - \partial_\nu A_\mu.
\end{eqnarray}
We may make $A_\mu$, $\phi$, and $x_\mu$ dimensionless by scaling by the appropriate power of $v$.  Furthermore, we make the rescaling scale $x^\mu \to x^\mu / e$ so that there is only one parameter in the model, $\beta$:
\begin{eqnarray}
{\cal D}_\mu \to e \left(\partial_\mu - i A_\mu\right) \\
\beta \equiv {\lambda \over 2 e^2}.
\end{eqnarray}

Our {\em anzatz} for the winding solutions will be
\begin{eqnarray}
\phi &=& f(r) e^{i n \theta} \\
A_\mu &=& (0,0,{n \over r} \alpha(r), 0).
\end{eqnarray}
We have used polar coordinates so that the components of $A_\mu$ are $(t,r,\theta,z)$.  The equations of motion which follow from this {\em anzatz} are
\begin{eqnarray}
0 &=& {1 \over r} \partial_r \left( r f'(r) \right)
- \left[ {n^2 \over r^2}(1-\alpha(r))^2 + \beta\left(f(r)^2 -1\right) \right] f(r) \\
0&=& r \partial_r \left({\alpha'(r) \over r}\right) - 2f(r)^2 (\alpha(r) -1).
\end{eqnarray}
We also need to specify boundary conditions.  To minimize potential and gradient terms at infinity, we must have
\begin{eqnarray}
\lim_{r \to \infty} \alpha(r) = 1, \qquad  \lim_{r \to \infty} f(r) = 1. \label{bc4Dfar}
\end{eqnarray}
With these values for $\alpha$ and $f$ far from the string, the fields will be gauge equivalent to the trivial winding vacuum: $A_\mu = 0$ and $\phi = 1$.
Continuity of the solutions in the core of the string imply
\begin{eqnarray}
\lim_{r \to 0} \alpha(r) = 0, \qquad  \lim_{r \to 0} f(r) = 0.  \label{bc4Dnear}
\end{eqnarray}
In fact, $\alpha$ must go to zero quickly (as $r^2$, not just as $r$) so as to avoid a divergence in the last term of the string tension (the gradient of the gauge field).  However, we cannot require this as a boundary condition since that would lead to an over-specified problem.  It will turn out, though, that this fall-off is true numerically.  For the string tension we find that
\begin{eqnarray}
\mu &=& 2 \pi v^2 \int r dr \left[ f'^2 + {n^2 \over r^2} (1-\alpha)^2 f^2
+ {\beta \over 2} (f^2-1)^2 + {n^2 \over 2 r^2} \alpha'^2 \right].
\end{eqnarray}
We may simplify this expression slightly by integrating the last term by parts and using the equation of motion for $\alpha$. 
\begin{eqnarray}
\mu &=& 2 \pi v^2 \int r dr \left[ f'^2 + {n^2 \over r^2} (1-\alpha) f^2
+ {\beta \over 2} (f^2-1)^2 \right] + 2\pi v^2 \lim_{r\to0} {\alpha \partial_r \alpha \over r} \\
&\equiv& 2 \pi v^2 \, g(\beta).
\end{eqnarray}
Notice that the factor $(1-\alpha)$ is no longer squared.  If we also make use of the fact that $\alpha$ goes to zero as $r^2$ then the last term is zero.  Following~\cite{VilenkinShellard} we have defined a new function $g(\beta)$ which we can numerically plot (see figure~\ref{gauge_tension_plot}) to understand the string tension as a function of our one free parameter.

\subsection{Local Symmetry with Brane Localized Higgs}
We now turn our attention to the $U(1)$ gauge symmetry in a flat extra dimension.  We will allow the gauge fields to propagate in the bulk, but restrict the Higgs to one boundary.  The action is
\begin{eqnarray}
S = \int d^4x\,dy \left\{ \left[ \left({\cal D}_{\mu}\phi \right)^* {\cal D}^{\mu}\phi
- {\lambda \over 4} \left( |\phi|^2 - v^2\right)^2 \right] \delta(y-L)
-{1 \over 4} F_{MN}F^{MN} \right\}
\end{eqnarray}
Now we must vary the action to find the equations of motion, taking care with the surface terms so that we also get the boundary conditions.  The bulk equation of motion for $A_M$ is simply
\begin{eqnarray}
\partial^M F_{MN} = 0, \label{5DgaugeEOM}
\end{eqnarray}
since in the bulk there is nothing but a free, Abelian, gauge field.  The scalar equation of motion has the same form as before since it is trapped on the 4-dimensional brane.  We repeat it here for completeness:
\begin{eqnarray}
0 = {\cal D}_\mu {\cal D}^{\mu} \phi 
+ {\lambda \over 2} \left( |\phi|^2 - v^2 \right) \phi.\label{5DscalarEOM}
\end{eqnarray}
This equation depends of course on the 5-dimensional parameters, such as the charge $e_5$, so that ${\cal D}_\mu = \partial_\mu - i e_5 A_\mu$.  The boundary conditions are:
\begin{eqnarray}
y=0:&&  F_{5 \mu} = 0 \nonumber \\
y=L:&&  F_{5 \mu} + ie_5 \left( \phi^* {\cal D}_\mu \phi - ({\cal D}_\mu \phi)^* \phi \right) =0.
\label{5dBC2}
\end{eqnarray}
First consider the vacuum, or trivial $n=0$ winding solution:
\begin{eqnarray}
\phi = v \qquad A_M=0
\end{eqnarray}
This solution to the equations of motion and the boundary conditions differs qualitatively from the global string case because there is no dependence on the extra dimension.  This would imply that a simple effective theory could capture all of the string physics if the winding solutions displayed the same homogeneity in the bulk.  However, we can see from the boundary condition that gradient terms from the winding of the Higgs will force the gauge field to have a profile in the bulk.  By looking ahead to equation~(\ref{EnergyDensity5d}) we may note that this vacuum solution has zero energy density, so we will not need to subtract off a (classical) cosmological constant as we did for the global string.

In order to be able to compare the winding solutions of this theory with an equivalent 4-dimensional theory, we need to consider equivalent parameters.  As we did for the global symmetry, we match the 4 and 5 dimensional parameters by considering the effective theory which results from substituting
\begin{equation}
A_M (x^\nu,z) \to A_\mu (x^\nu), \quad A_5=0
\end{equation}
and integrating out the extra dimension.  Because the scalar is trapped on the brane, $\lambda$ and $v$ will be equivalent to what we had before, which we have already anticipated by using the same variables.  However, the gauge field samples the bulk and a low energy observer scattering $\phi$ particles would measure a gauge coupling $e$ for an effective 4-dimensional theory given by
\begin{equation}
{ L \over e_5^2 } = { 1 \over e^2 }.
\end{equation}
Therefore we will want to express the problem in terms of $e$ (or $\beta$) rather than $e_5$.

We will look for static solutions which have $A_0 = 0 = A_5$.  We will again make everything dimensionless by scaling by the appropriate power of $v$.  We also rescale the four coordinates, $x^\mu$, by the 4-dimensional charge, set $\beta = \lambda/(2e^2)$ and rescale the gauge field by $\sqrt{L}$.  We scale $y$ by $L$ so that the new coordinate for the extra dimension runs from $0$ to $1$.  Note that this will not put all of the charge dependence in $\beta$ as it did for the 4-dimensional case.  It will still appear in the combination $L^2e^2 = L e_5^2$.  In the limit of a small extra dimension, this parameter will drop out when $e_5$ is taken to be fixed and we have just as many parameters as in the 4-dimensional case.

For a winding solution, we modify our {\em ansatz} by allowing the function $\alpha$ to have $y$ dependence:
\begin{eqnarray}
\phi &=& f(r) e^{i n \theta} \\
A_\mu &=& (0,0,{n \over r} \alpha(r,y), 0).
\end{eqnarray}
The equations of motion now become
\begin{eqnarray}
0 &=& \left[ L^2e^2 r \partial_r \left({1 \over r} \partial_r \right) + \partial_y^2 \right] \alpha(r,y), \label{gaugeEOM1} \\
0 &=& {1 \over r} \partial_r \left( r f'(r) \right)
- \left[ {n^2 \over r^2}(1-\alpha(r,y))^2 + \beta\left(f(r)^2 -1\right) \right] f(r). \label{gaugeEOM2}
\end{eqnarray}
The boundary conditions take the form
\begin{eqnarray}
y=0:&& \partial_y \alpha = 0 \label{gaugeBC1} \\
y=1:&& \partial_y \alpha = 2 L^2e^2 f^2(1-\alpha). \label{gaugeBC2}
\end{eqnarray}
Far from the string and in the core of the string $f$ and $\alpha$ must satisfy the same boundary conditions as the 4-dimensional winding solution, equations~(\ref{bc4Dfar}) and~(\ref{bc4Dnear}) respectively, so that the solution approaches a vacuum and is continuous.  The fact that the boundary condition far from the core of the string implies that the gauge field is independent of $y$ means that the total magnetic flux through the string is a constant in $y$ as can be determined by integrating the gauge field around the circle at infinity for any value of $y$.

The energy density is similar to the expression in the 4-dimensional case:
\begin{eqnarray}
\rho = {v^5 e^2 \over L} \left\{ \delta(y-1) \left(\left| {\cal D}_i \phi \right|^2 
+ {\beta \over 2} \left(|\phi|^2 - 1\right)^2 \right)
+ {1 \over 4} F^2_{ij} + {1 \over 2 L^2 e^2} F^2_{5j} \right\}, \label{EnergyDensity5d}
\end{eqnarray}
giving a string tension of
\begin{eqnarray}
\mu &=& 2\pi v^2 \int r dr \Bigg\{f'^2 + {n^2 \over r^2}(1-\alpha)^2f^2 + {\beta \over 2}(f^2-1)^2
\nonumber \\
&& \qquad \qquad \qquad + \int_0^1 dy {n^2 \over 2r^2}
\left( (\partial_r \alpha)^2 + {1 \over L^2e^2} (\partial_y \alpha)^2 \right) \Bigg\}.
\end{eqnarray}
As a check, we may notice that if $\alpha$ has no $y$ dependence, this expression reduces to exactly that of the string tension in the 4-dimensional case.  One might worry about what happens to the energy density as $L \to 0$, since $L$ appears in the denominator.  We can integrate the last term by parts, make use of the equations of motion and boundary conditions, and see that the string tension becomes
\begin{eqnarray}
\mu &=& 2 \pi v^2 \int r\,dr \left[ f'^2 + {n^2 \over r^2}f^2(1-\alpha) + {\beta \over 2}(f^2-1)^2 \right] \label{string_density} \\
&=& 2 \pi v^2 g(\beta),
\end{eqnarray}
where $\alpha$ is evaluated at $y=1$.  We have again made use of the fact that $\alpha$ is ${\cal O}(r^2)$ at small $r$.  This is exactly the same form as the expression for the 4-dimensional string tension.  Although the bulk contribution to the string tension drops out of this expression after using the equations of motion, there is still a nonzero energy density in the bulk, as can be checked with the numerical solutions below.

\subsection{Numerical Comparison}
Using a similar relaxational numerical technique as before to solve for the functions $f$ and $\alpha$ in both the 4 and 5-dimensional models we find that with reasonable initial guesses for both, they settle quickly into the sought-after solutions. The profile for the Higgs field, $f(r)$, is shown for both models in figure~\ref{higgs_profile}, while the profile of the gauge field, $\alpha$, is in figure~\ref{gauge_profile} and~\ref{gauge5d_profile}.  As can be seen from figure~\ref{gauge5d_profile}, the winding of the gauge field extends across the extra dimension without being exponentially damped as the scalar field profile was in the case of global symmetry breaking, figure~\ref{phi_winding_bulk_profile}. The gauge string is much more extended across the extra dimension than the global string was.
\DOUBLEFIGURE[t]{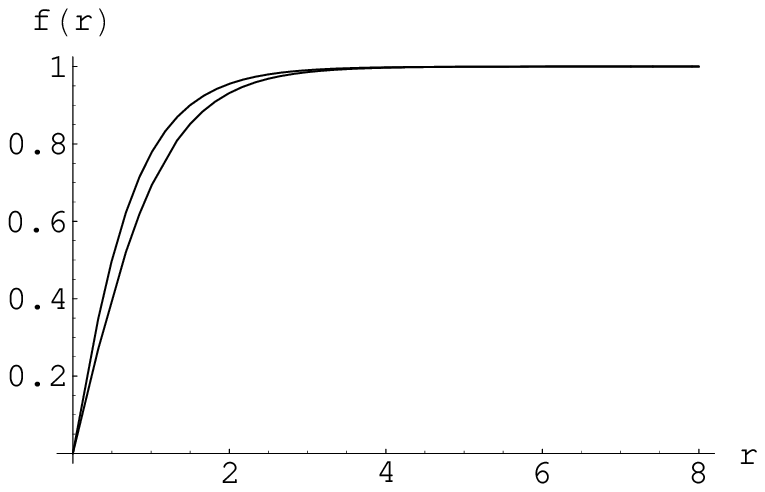,width=.4\textwidth}{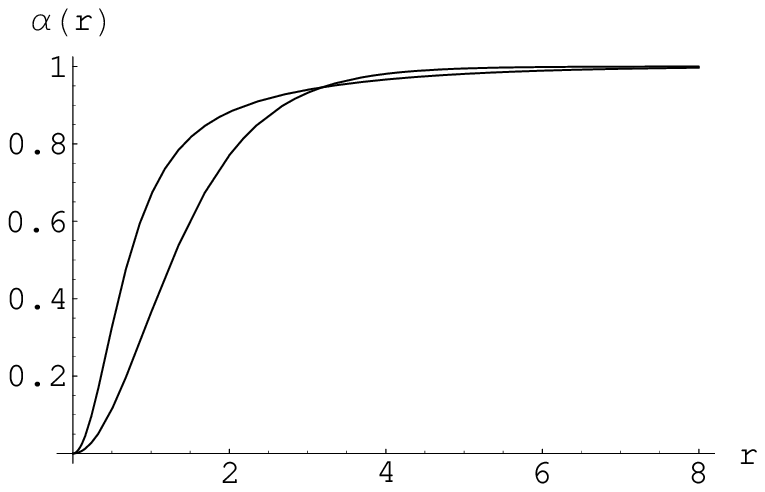,width=.4\textwidth}
{The profile for the Higgs field as represented by the function, $f(r)$.  The bottom curve is from the 4-dimensional model.  The parameters for this solution are $\beta = 1$ and for the 5-dimensional model, $e^2L^2=5$. \label{higgs_profile}}{The profile for the gauge field.  The bottom curve is $\alpha(r)$ from the 4-dimensional model while the top curve is the gauge field on the symmetry breaking brane: $\alpha(r,L)$.  Both curves are going to zero as $r^2$.  The model parameters are the same as before. \label{gauge_profile}}
\FIGURE[t]{\epsfig{file=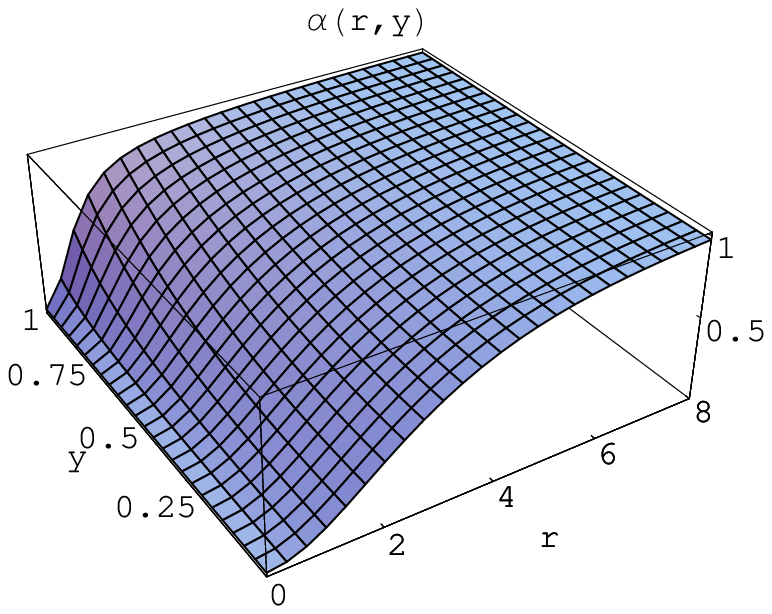,width=.6\textwidth}
\caption{The value of $\alpha(r,y)$ throughout the bulk.  The model parameters are the same as in figure~\ref{higgs_profile}.  The value of $r$ goes from $0$ to $8$, while $y$ goes from $0$ to $1$.
\label{gauge5d_profile}}}

We have calculated the energy density as a function of radius from the string, $r$, for each case and the result is shown in figure~\ref{gauge_tension_comparison}.  The energy density in the core of the 5-dimensional string is considerably higher than in the 4-dimensional string but this is more than offset in the string tension by the fact that the 4-dimensional string has more energy density near $r \sim 1$.  In both cases, the energy density falls off exponentially at large radius so that the total string tension is finite.

\FIGURE[t]{\epsfig{file=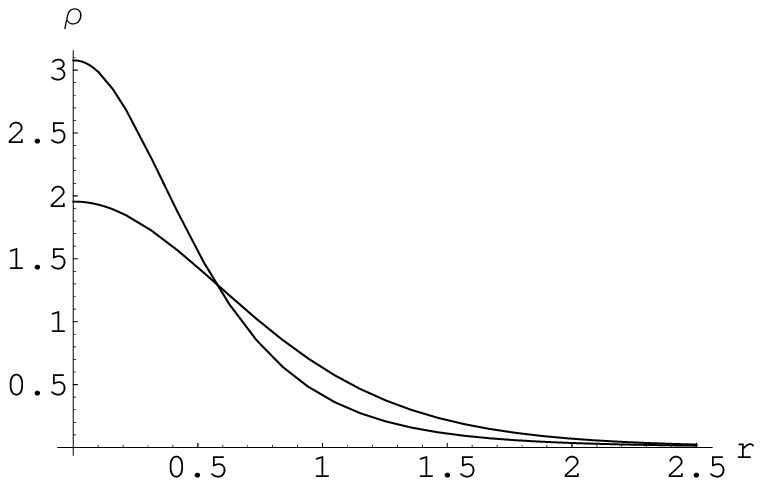,width=.6\textwidth}
\caption{The energy density of the string solution as a function of $r$.  This is the quantity in square brackets in equation~(\ref{string_density}) and this figure is analogous to figure~\ref{string_tension_comparison}.  The 5-d model peaks at a higher energy density in the core of the string, but the string tension comes from multiplying these functions by $2\pi r v^2$ and then integrating.  As a result, the string tension for the 5-d model is smaller.  The parameters are the same as figure~\ref{higgs_profile}. \label{gauge_tension_comparison}}}

\FIGURE[b]{\epsfig{file=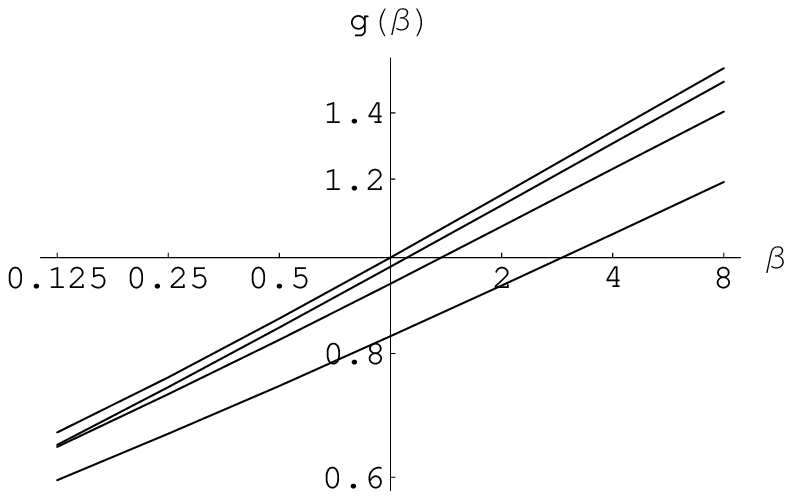,width=.6\textwidth}
\caption{The string tension, $g(\beta)$.  The top line is the 4-d model where $g(1)=1$.  The next three lines down are for the 5-d model with $e^2L^2 = 0.2,1.0,5.0$ respectively. \label{gauge_tension_plot}}}

We have also plotted the string tension, $g(\beta)$, for the 4-dimensional model and for various values of $e^2L^2$ in the 5-dimensional model, figure~\ref{gauge_tension_plot}.  The string tension of the 5-dimensional model seems to be converging to the 4-dimensional model as $e^2L^2$ becomes small, as we expect.

\section{Higgsless Models}\label{HiggslessSection}
Some recent efforts to address the hierarchy problem with an extra dimensional setup have made use of boundary conditions, rather than a Higgs field, to break the symmetry spontaneously~\cite{Higgsless}.  It was found that in a pure gauge theory, the variation of the action could be set to zero through several different choices of boundary conditions.  In a gauge theory the choice of Dirichlet conditions on one end and Neumann on the other, for example, had the effect of giving a mass to the lowest KK mode for the gauge field which would otherwise have been massless.  That this gauge symmetry breaking is spontaneous might be guessed from the fact that the Lagrangian is gauge invariant.  In addition the unitarity violations which are common to generic massive gauge theories do not occur if the mass arises from boundary conditions through this action principle.

We want to understand then what happens to topological defects if the symmetry is broken by boundary conditions.  Without a potential for the Higgs boson with a non-trivial vacuum manifold we might immediately think that there is no possibility for a stable defect solution.  However, in the Higgsless construction of gauge symmetry breaking, the component of the gauge field in the compact dimension, $A_5$, plays the role of the Goldstone boson which is eaten to make the gauge field massive.  It seems that there is some possibility for structure in the $A_5$ field to give rise to a winding.  We will see, though, that for an Abelian theory this does not happen.  The structure of the equations of motions and boundary conditions allow for a very simple scaling argument which rules out any stable, finite energy, defects in an Abelian Higgsless model.  In non-Abelian theories, the picture is more complicated and there may still be a possibility for defects in some cases.

At this point, the choice of boundary conditions seems like an arbitrary choice for the model builder to make.  However, in other contexts, such as a vibrating guitar string, boundary conditions are imposed by knowing the physics of the boundaries.  The situation is really no different here and there should be a more fundamental theory which explains the choice of boundary conditions.  The case of Dirichlet boundary conditions, $A_\mu = 0$, arise from a Higgs field where the VEV is taken to infinity, while Neumann conditions, $\partial_5 A_\mu = 0$, arise if there is no breaking of the gauge theory on the boundary.

Examples of these results for the boundary conditions can be seen in equations~(\ref{5dBC2}) for the Abelian model we have already considered.  By working in the $A_5=0$ gauge we see that the large VEV limit of $\phi$ imposes Dirichlet boundary conditions on the symmetry breaking brane at $y=L$.  For this Abelian model then, the equations of motions remain:
\begin{equation}
\partial^M F_{MN} = 0,
\end{equation}
while the boundary conditions have become
\begin{equation}
\left . F_{\mu5}\right |_0 = 0, \quad \left . A_\mu \right |_L = 0,
\end{equation}
which are all linear in $A_M$ and homogeneous.  If $A_M$ is a solution with string tension $\mu$, then $\lambda A_M$ is also a solution with tension $\lambda^2 \mu$ since the tension is quadratic in the gauge field.  Intuitively then it is clear that any hypothetical solution, $A_M$, is smoothly connected to the vacuum solution with zero tension and $A_M=0$.  There are, therefore, no stable defects with only a gauge field.  We will make this statement more precise below and show that this is a special case of the scaling arguments employed by Derrick's Theorem~\cite{VilenkinShellard}.

First, though, we consider the large VEV limit of the winding solution we already have for the local $U(1)$ breaking with a Higgs field.  Recall that we scaled all dimensionful parameters by $v$ to make the system dimensionless and as a result, the parameter $v$ made an appearance only in the string tension.  Therefore, the tension will diverge as $\mu \sim v^2$, but the profile of the winding will remain constant in the variables used in equations~(\ref{gaugeEOM1}) to~(\ref{gaugeBC2}).  In physical coordinates then, the size of the string will shrink in the radial direction, $r \sim 1/v$.  In the fifth dimension the profile of the string will remain constant since we worked with a variable which was also scaled by the size of the extra dimension, $(yv)/(Lv)$, which is constant as $v$ grows.  We can now see that the Higgsless limit of a broken $U(1)$ gauge theory may contain infinitely massive and thin strings, and that no new windings make an appearance.

\subsection{Higgsless Equations of Motion and Boundary Conditions}
We now consider a non-Abelian Higgsless gauge theory.  In this subsection we review the conditions at the boundaries so that the action is minimized.  Consider a generic non-Abelian gauge theory with no scalars:
\begin{equation}
S = {1 \over 4} \int_0^L dy \int d^4x \, F^a_{MN} F^{aMN}, \label{GaugeAction}
\end{equation}
where the field strength tensor is defined as usual,
\begin{equation}
F^a_{MN} = \partial_M A^a_N - \partial_N A^a_M + g f^{abc} A^b_M A^c_N.
\end{equation}
Let us vary the action, taking care with any integration by parts and the boundary terms.  We get
\begin{eqnarray}
\delta S &=& \int_0^L dy \int d^4x \Big\{\delta A^a_{\nu}\left(\delta^{ac}\partial_M+g f^{abc}A^b_M \right)
F^{c\nu M} \nonumber \\
&& \qquad\qquad\qquad\, - \delta A^a_5 \left(\delta^{ac}\partial_{\mu}+g f^{abc}A^b_{\mu}\right)
F^{c\mu 5} \Big\} \nonumber \\
&& + \int d^4x \, \delta A^{a\mu}F^a_{\mu 5} \big|^{y=L}_{y=0}
\end{eqnarray}
Requiring that the variation vanish implies the bulk equation of motion:
\begin{equation}
\left( \delta^{ac} \partial_M + g f^{abc} A^b_M \right) F^{cNM} = 0, \label{bulk_gauge_EoM}
\end{equation}
together with the boundary conditions
\begin{equation}
F^a_{\mu 5} = 0 \quad \textstyle{or} \quad A^a_\mu = 0.
\end{equation}
We have used the fact that the metric is mostly minus to raise or lower the $5$.

By varying the action for our gauge theory with respect to the metric, we find that the stress-energy tensor is
\begin{equation}
T_{\alpha \beta} = {1 \over 4} \eta_{\alpha\beta} F^a_{\mu\nu} F^{a\mu\nu}
- F^{a\mu}_\alpha F^a_{\beta\mu},
\end{equation}
and the energy density is always positive:
\begin{equation}
T_{00} = {1 \over 2} \sum_{a,i} \left(F^a_{0i}\right)^2 + {1 \over 4} \sum_{a,i,j} \left(F^a_{ij}\right)^2.
\label{gauge_energy_density}
\end{equation}
Now notice that $A^a_{\mu}(x^\nu,y) \equiv 0$ is a solution to the bulk equation of motion as well as to either of the possible boundary conditions.  Thus, the trivial solution is the lowest energy solution.  This differs from the case of the scalar field where the $\phi=0$ solution had higher energy than the $n=0$ solution.

\subsection{Generalized Derrick's Theorem}
By considering a generalization of Derrick's Theorem (see ref.~\cite{VilenkinShellard}) in spaces with compact extra dimensions we will find that our options for finite energy, time-independent defects are quite limited.  First, however, we reconsider a simple scaling argument to make sure we understand what happens in finite dimensions.  Suppose we have an integral over $m$ dimensions of a derivative of some function, for example:
\begin{equation}
I \equiv \int d^m x \vec{\nabla}_x^2 f\left(\vec{x}\right),
\end{equation}
where we have not yet specified whether these dimensions are compact so the limits have not yet been written in.  The function $f\left(\vec{x}\right)$ represents some field configuration which we can rescale by replacing it with $f\left(\lambda\vec{x}\right)$.  For $\lambda > 1$ we are shrinking the configuration: think of $f\left(\lambda\vec{x}\right) = \exp\left(-\lambda^2\vec{x}^2\right)$.  Now the family of integrals generated by this family of field configurations is given by
\begin{equation}
I(\lambda) \equiv \int d^m x \vec{\nabla}_x^2 f\left(\lambda\vec{x}\right),
\end{equation}
and can be related to the original integral $I$ by substituting
\begin{eqnarray}
\vec{z} \equiv \lambda \vec{x}, \\
d^m x = \lambda^{-m} d^m z, \\
\partial_x = \lambda \partial_z.
\end{eqnarray}
After making this substitution we have
\begin{equation}
I(\lambda) = \lambda^{2-m} \int d^m z \vec{\nabla}_z^2 f\left(\vec{z}\right),
\end{equation}
but we must now consider the limits of integration.  If the dimensions are infinite and the integral runs over all space then the limits are unchanged by this rescaling of $\vec{x}$.  This leaves:
\begin{equation}
I(\lambda) = \lambda^{2-m} I.
\end{equation}
However, if the limits in $x$ run from $a$ to $b$ (where these are representing multidimensional quantities) then all we can say is:
\begin{equation}
I(\lambda) = \lambda^{2-m} \int_{\lambda a}^{\lambda b} d^m z \vec{\nabla}_z^2 f\left(\vec{z}\right),
\end{equation}
and we are left with no simple scaling relation for the integral.  Therefore, in what follows, we will only look for scaling arguments in the infinite dimensions.  Of course, if we have a large compact dimension which is larger than the length scales of the field configuration then we may be able to treat those dimensions as infinite so long as changing the limits of integration has no impact on the integral.  By compact, we mean then, any dimension which is of the size of, or smaller than, the field configuration of interest.

In an important aside, if the integral $I$ represents an action or an energy which is {\em infinite} then we must regulate by cutting off the spatial integral.  We now see that this will ruin the scaling argument.  This is why Derrick's theorem does not rule out the global string which has a divergent Lagrangian in an infinite space.

Consider now $1+d+n$ dimensions where $d$ spatial dimensions have infinite extent and $n$ spatial dimensions have finite extent.  The action for a non-Abelian theory with only gauge fields is:
\begin{equation}
S = {-1 \over 4} \int dt \int d^d x \int d^n y F_{MN}^a F^{a MN},
\end{equation}
so that the Lagrangian can be written as
\begin{eqnarray}
L &=& {1 \over 4} \int d^d x \int d^n y \Big[ 2 \left(F^a_{0 x_i}\right)^2  
+  2 \left(F^a_{0 y_i}\right)^2 \nonumber \\
&& \qquad - \left(F^a_{x_i x_j}\right)^2 - \left(F^a_{y_i y_j}\right)^2 
- 2 \left(F^a_{x_i y_j}\right)^2\Big] \\
&\equiv& L_1 + L_2 - L_3 - L_4 - L_5,
\end{eqnarray}
where sums over the gauge and dimensional indices are assumed and we are using a mostly minus metric.  Each of the $L_i$ are non-negative.  Furthermore, we may calculate the energy of a static configuration to be the sum of these five terms:
\begin{equation}
E = \int d^d x \int d^n y \, \rho = \sum_{i=1}^5 L_i.
\end{equation}
If we seek a finite energy solution, then each of the $L_i$ must be finite.

Let us first consider an Abelian theory and the scaling of the different terms in the Lagrangian with three parameters, $\lambda$, $\sigma$, and $\gamma$ as follows
\begin{eqnarray}
\sigma \lambda A_0 \left(\lambda \gamma \vec{x}, \vec{y} \right) \\
\lambda \gamma^2 A_{x_i} \left(\lambda \gamma \vec{x}, \vec{y} \right) \\
\gamma A_{y_i} \left(\lambda \gamma \vec{x}, \vec{y} \right).
\end{eqnarray}
Notice that we are looking for time-independent solutions here.  With this rescaling of the gauge field configuration the Lagrangian scales as
\begin{equation}
L \to \lambda^{-d} \gamma^{-d} \left[ \sigma^2 \left( \lambda^4 \gamma^2 L_1 + \lambda^2 L_2 \right)
- \lambda^4 \gamma^6 L_3 - \gamma^2 L_4 - \lambda^2 \gamma^4 L_5 \right].
\end{equation}
Since we seek a solution to the equations of motion, any solution must be stationary at $\sigma = 1 = \lambda$.  By varying $\sigma$ we see immediately that $L_1 = 0 = L_2$.  By varying $\lambda$ and $\gamma$, we get two equations for the remaining parts of the Lagrangian.  Subtracting these two equations gives, $L_3+L_4+L_5=0$ and we see that they must all vanish (since they are individually non-negative).  This seems to preclude the existence of defects.  However, it really only rules out time-independent, finite energy solutions, not time-dependent, non-dissipative or infinite energy solutions.

For a non-Abelian theory, the scaling with $\gamma$ will transform the $L_5$ term inhomogeneously:
\begin{equation}
F_{x_i y_j} \to \lambda \gamma^2 \left( \partial_{x_i} A^a_{y_j} - \partial_{y_j} A^a_{x_i} \right)
+ \lambda \gamma^3 g f^{abc} A^b_{x_i} A^c_{y_j}.
\end{equation}
We can therefore only make use of the scaling with $\sigma$ and $\lambda$.  We are left with the constraint:
\begin{equation}
(4-d)L_3 - d L_4 + (2-d) L_5 = 0.
\end{equation}
It seems then that static defect solutions may be possible for non-Abelian theories so long as $d$ is less than or equal to $4$.  It would certainly be interesting to find an example of a new defect solution in a Higgsless model, but we leave this for future work.

\section{Summary}\label{SummarySection}
We have considered several scenarios of extra dimensions with localized symmetry breaking.  In the first, the breaking of a global $U(1)$ symmetry by boundary terms forced the scalar to acquire a vacuum expectation value (VEV).  This system was simple enough that many features of the possible static, classical, solutions could be understood qualitatively.  The magnitude of the VEV could be understood via an effective theory where an integration was performed over the extra dimension.  However, in the case of windings, the tension in the resulting string was less than we would have expected from the effective theory.  This happens because the winding forces the profile of the scalar in the new dimension to differ from what it would have been without the winding.  In some sense, the energy density in the core of the string is high enough that the string is sensitive to the full 5-dimensional theory.

The gauge string retained many similarities with the global string.  For example, the tension of the solution from the full 5-dimension theory was less than that for the equivalent 4-dimensional effective theory.  Due to the finite nature of the tension, in this case, we were able to make more meaningful quantitative comparisons between the 4 and 5-dimensional theories.  A significant qualitative difference is that the gauge string has a winding which is nearly constant throughout the bulk of the extra dimension, while the global string was exponentially localized on the symmetry breaking brane.

In both cases, the strings are not going to be able to miss each other in the extra dimension since they are both required to have winding of the scalar on the symmetry breaking brane and some amount of the winding is carried across the entire bulk of the extra dimension.  In this sense they are not genuine strings and if the symmetry breaking terms were homogeneous throughout the extra dimension they would be 2-branes.  It remains an open question as to whether or not the structure of the strings in the extra dimension changes the dependence of the inter-commutation probability on the string velocity.

Finally, we looked at the Higgsless limit of the local symmetry breaking and found that the string tension scales with the square of the VEV of the scalar and the radius of the string scales with the inverse of the VEV leading to massive, thin strings.  Through a generalization of Derrick's theorem we showed that there are no new static, finite energy, winding structures possible for an Abelian theory.  This may provide another way to remove unwanted topological defects from a theory.  A non-Abelian theory, however, may be able to support windings without a Higgs field, but we have not yet investigated this possibility in depth.

There are a number of interesting follow-up questions we can ask at this point. First, what happens when the internal space is warped, such as in the Randall-Sundrum models \cite{randallsundrum}? We would expect that the existence of a new length scale such as the AdS curvature scale would modify both the structure of defects as well as our discussion of how well the effective field theory obtained by integrating out the warped extra dimension would account for the energy density of the defect. 

A second question, already alluded to above is what happens to baryogenesis in theories with localized symmetry breaking or in the Higgsless case. If the energy of the sphaleron is modified, this will have a profound effect on the rate of baryon number violation, since the energy appears in the Boltzmann exponential. In the Higgsless case, we need to examine the theory to see if anything like the sphaleron even exists!

There are other topics of interest; changes in the evolution of cosmic string networks, for example, but we hope that we have persuaded the reader that this line of research will yield fruitful new insights into many interesting phenomena.

\acknowledgments
MM would like to thank Alex Friedland for valuable discussions and encouragement on this project.  RH is supported in part by DOE grant DE-FG03-91-ER40682.  MM is supported in part by the DOE under contract W-7405-ENG-36.

\end{document}